\documentstyle[aps,prb,preprint]{revtex}
\begin{document} 
\draft 
\title{ Phase diagram of the restricted solid-on-solid model coupled 
        to the Ising model}
\author{Sooyeul Lee and Koo-Chul Lee} 
\address{Department of Physics and Center for Theoretical Physics\\
     Seoul National University\\
     Seoul 151-742, Korea} 
\author{J. M. Kosterlitz}
\address{Department of Physics\\
         Brown University\\
         Providence, Rhode Island 02912}
\maketitle 
 
\begin{abstract} 
We study the phase transitions of a restricted solid-on-solid model
coupled to an Ising model, which can be derived from the coupled $XY$-Ising
model. There are two kinds of phase transition lines. One is a Ising transition
line and the other is surface roughening transition line. The latter is a KT
transition line from the viewpoint of the $XY$ model.
Using a microcanonical Monte Carlo technique, we  obtain a
very accurate two dimensional phase diagram. 
The two transition lines are separate in all the parameter space
we study. This result is strong evidence that the fully
frustrated $XY$ model orders by two separate transitions and that roughening
and reconstruction transitions of crystal surfaces occur separately.

\end{abstract} 
\narrowtext 
 
\section{Introduction}\label{sec1} 

Since the pioneering work of Teitel and Jayaprakash~\cite{teitel}, two
dimensional (2D) fully frustrated $XY$ (FF$XY$) models have been intensively 
studied over the last decade, both analytically and 
numerically.~\cite{mychoi,grest,jrlee,jlee,slee,olsson,thijssen} This model
has a continuous $U(1)$ symmetry associated with global spin rotations and
also a discrete $Z(2)$ symmetry because of the double degeneracy of the
ground state vortex configurations. At sufficiently low temperature $(T)$
the system will be completely ordered with true long range order in the
Ising order parameter characterizing the vortex lattice and algebraic decay
of correlations of the $XY$ order parameter. As $T$ is increased, the system
will disorder by one of two possible scenarios. One is the simultaneous loss
of both $XY$ and Ising order and the other is a two stage process in which
$XY$ order is destroyed at a lower temperature than the Ising order. Since
the original paper on this model,~\cite{teitel} an enormous effort has gone
into trying to clarify the nature of the phase transition(s) of this model
and several related models believed to be in the same universality class
have been studied numerically.~\cite{grest,jrlee,jlee,slee,olsson,thijssen}
However, the nature of the transition(s) in the FF$XY$ model and its
generalizations is still somewhat controversial. Monte Carlo (MC)
simulations of the Coulomb gas with half integer charges~\cite{grest,jrlee}
generally show a vortex lattice melting transition and a KT-like transition
at slightly different temperatures which is consistent with the second
scenario. However, the melting transition is expected to be in the Ising
universality class but the simulations yield exponents which are
inconsistent with the Ising values. Moreover, at the KT-like transition the
discontinuity in the helicity modulus is much larger than the universal
value of $2/\pi$~\cite{nk}. Yet another model is the coupled $XY$-Ising
model~\cite{jlee} which is a coarse-grained version of the FF$XY$ model. It
was suggested that this model displayed both possible scenarios depending on
the parameters of the model. Recent studies~\cite{slee,olsson} of the FF$XY$
model
with a nearest neighbor cosine interaction show that the disordering is by
the second scenario with a KT transition followed by an Ising transition at a
slightly higher temperature. From a theoretical point of view, the FF$XY$
model may have separate $XY$ and Ising transitions if the critical exponents
of the vortex lattice melting are consistent with the 2D Ising values and
the jump in the helicity modulus at the $XY$ transition has the universal
value of $2/\pi$ but most numerical studies are not consistent with these.
However, in his recent numerical study, Olsson~\cite{olsson} pointed out
that, for conventional finite size scaling to hold, it is essential that the
system size $L$ is much larger than any other length scale. Since the vortex
lattice melting temperature $T_{c\rm{I}}$ is only slightly larger than $T_{\rm{KT}}$,
the relevant length is $\xi_{+}(T_{c\rm{I}})$, the $XY$ correlation length at the
lattice melting temperature. As is well known, this remains rather large
quite far above $T_{\rm{KT}}$ and, by taking this into account, a good fit to the
double transition scenario was obtained~\cite{olsson} with a conventional KT
transition followed by an Ising transition.

Whatever the precise nature of the transitions in any model, an essential
first step is to determine whether the system has two separate transitions
or a single transition. It is of some interest from the point of view of
conformal field theories to find a non-trivial model in which the central
charge $c>1$. Recently, a related class of models for transitions on crystal
surfaces with competing reconstruction (Ising) and roughening ($XY$) degrees
of freedom has been studied~\cite{nijs,carlon}.
Den Nijs has studied competition between surface
roughening and reconstruction and tentatively concluded that the roughening
and deconstruction transition lines meet and finally merge together into 
a single roughening induced simultaneous reconstruction transition
line.~\cite{nijs} On the other hand, Carlon in a recent study of the
staggered six vertex model,~\cite{carlon}
has suggested that the two transition lines never merge but become very
close to each other which is strong
evidence for two separate transition in the FF$XY$ model. In order to
clarify whether the Ising and $XY$ transition lines merge we introduce the 
restricted solid-on-solid (RSOS) model coupled to the Ising model. This 
RSOS-Ising model can be derived from  the $XY$-Ising
model by making a duality transformation on the $XY$ degrees of freedom which
introduces integer height variables on the dual lattice.
The resulting RSOS-Ising model has both Ising and height ($XY$) degrees of
freedom and, for this reason, can be considered to have the same features
as the staggered six vertex model and reconstructed surface models.

We investigate the RSOS-Ising system
by a rejection free microcanonical MC technique~\cite{kclee} which is
appropriate to this system and also highly accurate. We have
obtained an accurate phase diagram in a two parameter space
and some critical exponents. As a result we have found there are
Ising and roughening critical lines which never merge in all
parameter space. This result is strong evidence against the single
transition scenario in the fully frustrated $XY$ model.

In the following section we give a derivation of the RSOS-Ising model from
the coupled $XY$-Ising system. In section III we present a brief explanation of
the MC procedure~\cite{kclee} as adapted for this system, methods of data
analysis and MC results. Finally we present our conclusions.

\section{Restricted Solid-on-Solid model coupled to Ising model}

From renormalization group and universality arguments the isotropic FF$XY$
model can be described by a coupled $XY$-Ising model of the form
\begin{equation}
   -\beta H = A_0 \sum_{\left < i,j \right>} (1+\sigma_i\sigma_j)
                  \cos(\theta_i-\theta_j) + 
              C_0 \sum_{\left < i,j \right>} \sigma_i\sigma_j,
\label{0}
\end{equation}
where $\theta_i$ is the phase of the continuous spin at $i$th site and 
$\sigma_i=\pm 1$ describes the chirality. The physical range of the
phenomenological parameters is $A_{0}>0$ and $A_{0}+C_{0}>0$.
This model has been intensively studied by Lee {\it et al.}~\cite{jlee}
By studying the behavior of the chiral (Ising) degrees of freedom
$\sigma_{i}$, they found a line of continuous transitions with a segment of
non-Ising like critical exponents, in particular $\nu_{I}\approx 0.8<1$ and
$\eta_{I}>1/4$. Based on this, they argued that the Ising and $XY$ critical
lines must have merged into a single line of simultaneous ordering of both
degrees of freedom. However, they made no explicit study of the $XY$ degree
of freedom for technical reasons, which weakens their argument. To
simultaneously study the $XY$ and Ising critical lines, we make a duality
transformation~\cite{jkkn} on the $XY$ variables of Eq.(\ref{0}) to obtain a
restricted solid-on-solid model coupled to an Ising model, in which only
discrete variables are involved, as described below.

The partition function of the $XY$-Ising model of Eq.(\ref{0}) reads
\begin{equation}
 Z=\sum_{\{\sigma\}} \exp \left [ C_0\sum_{ \left < i,j \right > } 
                                     \sigma_i\sigma_j \right ]
  \int_0^{2\pi}\cdot\cdot\cdot \int_0^{2\pi} \prod_{k=1}^N {\mbox d}\theta_k
   \prod_{\left <i,j\right >} \exp \left [ K_{ij} \cos(\theta_i-\theta_j)
                                      \right ],
\end{equation}
where $\{\sigma\}$ denotes the sum over all configurations of Ising spins,
$N$ is the number of lattice sites and $K_{ij}=A_0(1+\sigma_i\sigma_j)$.
Using the periodicity of the integrand in 
$\theta_{ij}\equiv \theta_i-\theta_j$ we have
\begin{equation}
  \exp \left [ K_{ij} \cos(\theta_i-\theta_j) \right ] =
  \sum_{n_{ij}=-\infty}^{\infty} \exp \left [ in_{ij}\theta_{ij} \right ]
                                 \exp \left [ \tilde{V}(n_{ij}) \right ],
\end{equation} 
where $\exp \left [ \tilde{V}(n_{ij}) \right ]$ is a Fourier component of 
$\exp \left [ K_{ij} \cos(\theta_i-\theta_j) \right ]$. This yields
\begin{equation}
 Z=\sum_{\{\sigma\}} \exp \left [ C_0\sum_{ \left < i,j \right > } 
                                     \sigma_i\sigma_j \right ]
   \int_0^{2\pi}\cdot\cdot\cdot\int_0^{2\pi} \prod_{k=1}^N {\mbox d}\theta_k
   \prod_{\left < i,j \right >} \sum_{n_{ij}=-\infty}^\infty
   \exp \left [ in_{ij}\theta_{ij} + \tilde{V}(n_{ij})\right ]. 
\end{equation}
The integrations over the $\theta_{k}$ can be easily done and, noting that
$n_{ij}=-n_{ji}$, give $N$ constraints on the $2N$ integers $n_{ij}$ so the
partition function has the form
\begin{equation}
 Z=\sum_{\{\sigma\}} \exp \left [ C_0\sum_{ \left < i,j \right > } 
                                     \sigma_i\sigma_j \right ]
   \sum_{\{n\}}\left [ \prod_i \Delta(i) \right ]
   \prod_{\left <k,l \right>} \exp \left [ \tilde{V}(n_{kl})\right ]
\end{equation} 
where the $\Delta (i)$ denote the constraints at site $i$.
For example, the case as shown in Fig.~1 gives
$\Delta(0) = \delta_{n_{01}+n_{02}, n_{30}+n_{40}}$.
The $N$ constraints are automatically satisfied by introducing the
integer height variables $h_{I}$ on the dual lattice shown in Fig.~1 as
\begin{eqnarray*}
  n_{01} &=& h_1 - h_2, \ \ n_{02} = h_2 - h_3 \\
  n_{30} &=& h_4 - h_3, \ \ n_{40} = h_1 - h_4.
\end{eqnarray*} 
In terms of the heights $h_{I}$ the partition function has the apparently
simple form
\begin{equation}
\label{eq.6}
 Z=\sum_{\{\sigma\}} \exp \left [ C_0\sum_{ \left < i,j \right > } 
                                     \sigma_i\sigma_j \right ]
   \sum_{\{h\}} \exp \left [ \sum_{\left < I,J \right >} 
				     \tilde{V}(h_I-h_J)\right ]
\end{equation}  
where $I$ and $J$ are dual lattice sites and $-\infty<h_{I}<\infty$. To this
point everything is exact but Eq.(\ref{eq.6}) is not suitable for simulation
because of the infinite range of integer height differences $h_{I}-h_{J}$.
The Fourier component is
\begin{equation}
 \exp \left [\tilde{V}(h_I-h_J)\right ] 
 = \frac{1}{2\pi} \int_0^{2\pi}
      \exp \left [-i(h_I-h_J)\theta_{ij} \right ]
      \exp \left [ A_0(1+\sigma_i\sigma_j)\cos \theta_{ij} \right ]
   {\mbox d} \theta_{ij},
\label{4}
\end{equation}
where the nearest neighbor bonds $IJ$ on the dual lattice and $ij$ on the
original lattice cross each other. If $\sigma_i\sigma_j=-1$, the Fourier
component $\exp \left [\tilde{V}(h)\right ]=\delta_{h,0}$. A step in $h_{I}$
cannot cross an Ising domain wall. When $\sigma_i\sigma_j=1$
Eq.~(\ref{4}) becomes $I_h(2A_0)$, where $I_h$ is a Bessel function of order
$h$ with imaginary argument and $h=h_I-h_J$ and $I_h(0)=\delta_{h,0}$.
From the asymptotic behavior of the Bessel functions we have
$I_h(2A_0)/I_0(2A_0) \approx \exp(-h^2/4A_0)$, which decays rapidly as
$h$ increases if the inverse temperature $A_0$ is not too large. Therefore,
it is a reasonable approximation to keep only the contributions from
$h_{I}-h_{J}=0,\pm 1$ when $\sigma_i \sigma_j = 1$.

In view of these arguments, an effective Hamiltonian for the system is
\begin{equation}
 -\beta H = C\sum_{\left <i,j\right > } \sigma_i\sigma_j -
	    A\sum_{\left <I,J\right > } \left | h_I-h_J \right |
\label{1}
\end{equation}
where $(h_{I}-h_{J})=0,\pm 1$ if the corresponding
$\sigma_i\sigma_j=1$ and $h_I=h_J$ otherwise.
By comparing the Boltzmann weights derived from the Hamiltonians (\ref{0})
and (\ref{1}) we obtain the relations
$\exp (-A) = I_{1}(2A_{0})/I_{0}(2A_{0}) $ and 
$\exp(2C) = \exp(2C_{0}) I_{0}(2A_{0})$. $A$ is inversely
related to $A_{0}$ and $C$ is a linear sum of $C_{0}$ and $A_{0}$.
Eq.~(\ref{1}) is the Hamiltonian for a RSOS-Ising model and the heights
$h_I$ and Ising spins $\sigma_{i}$ are coupled via the constraint that a
step in $h$ is forbidden to cross a domain wall in $\sigma$.

\section{Monte Carlo Analysis}

Due to the restrictions on the spin and height variables discussed in the
last section, some spin-height configurations are not allowed which we
call forbidden configurations.
In our MC procedure, we select a spin or height variable randomly and
update the configuration by a trial move.
The trial move for the spin variable $\sigma$ is a simple flipping, $i.e.$,
$\sigma \rightarrow -\sigma$.
For the height variable  $h$, the trial move is either raise or lower the selected height by $1$ unit, $i.e.$,
$h \rightarrow h \pm 1$.
If the attempted move takes the system to a forbidden configuration,
we call the move forbidden and otherwise we call the move allowed.
The number of forbidden moves changes
according to the configuration of the neighboring spins and heights.
For accuracy and efficiency, we use the microcanonical MC method~\cite{kclee}
introduced by one of the authors to study
thermodynamic functions of the system in the 2D parameter
space $(A,C)$. We further apply a rejection-free technique to speed up the
MC procedure.~\cite{kclee} To apply the microcanonical MC
method to the RSOS-Ising model we first introduce an energy-like quantity
\begin{equation}
  E = E_{\rm {I}} + E_{\rm {R}},
\end{equation}
where $E_{\rm {I}} = -\sum_{\left < i,j \right >} \sigma_i\sigma_j$ and
$E_{\rm {R}}=2\sum_{\left <I,J\right >} \left |h_I-h_J\right|$.
 The factor 2 in $E_{\rm {R}}$
is introduced for convenience to make the minimum energy changes
$\Delta E_{\rm {R}}=\Delta E_{\rm {I}}=4$ for both $E_{\rm {R}}$ and
$E_{\rm{I}}$. The quantities to be computed are the
number of configurations $\tilde{\omega}(E,E_{\rm{I}})$ and the microcanonical
average $\tilde{Q}(E,E_{\rm{I}})$ of a thermodynamic quantity $Q$ at a set of
values of $E$ and $E_{\rm{I}}$.

Since the rejection free MC technique~\cite{kclee} is not well known, we
present a brief account of the microcanonical MC method as adapted
to the RSOS-Ising system. We set up a random walk through a configuration
space restricted to a small energy square in the $(E, E_{\rm{I}})$ domain given by
\begin{eqnarray}
E^{(i)} - q\Delta E /2&\leq& E_{\phantom{\rm{I}}}
		\leq E^{(i+1)} + q\Delta E /2
\label{110} \\
E_{\rm{I}}^{(j)} - q\Delta E_{\rm {I}}/2 &\leq& E_{\rm{I}}
		\leq E_{\rm{I}}^{(j+1)} + q\Delta E_{\rm {I}}/2
\label{111}
\end{eqnarray}
where $\Delta E=4$ is the minimum energy separation in $E$, $q$ is the
coordination number of the lattice,
$E^{(i+1)} = E^{(i)} + \Delta E$ and
$E_{\rm{I}}^{(j+1)} = E_{\rm{I}}^{(j)} + \Delta E_{\rm {I}}$.
$q\Delta E/2$ and $q\Delta E_{\rm {I}}/2$ are
the maximum energy changes in $E$ and  $E_{\rm {I}}$ when a spin or
height variable is updated to a new state. The size of the square given by
Eqs.~(\ref{110}) and (\ref{111}) is chosen to make  any allowed move
keep the new configuration within this square  if the current
configuration is in the innermost square, a configuration whose
$E$ and  $E_{\rm {I}}$ values belonging to one of four sets,
$(E^{(i)},E_{\rm {I}}^{(j)})$, $(E^{(i+1)},E_{\rm {I}}^{(j)})$,
$(E^{(i)},E_{\rm {I}}^{(j+1)})$, and $(E^{(i+1)},E_{\rm {I}}^{(j+1)})$.

Since there are $N$ candidates for the trial spin move and $2N$
candidates for the trial height move, there is a total of
$\tilde{N} \equiv 3N$ of candidates for the trial move.
Then out of this $\tilde{N}$, there are
$\tilde{N}_{\rm{alwd}}$ allowed and
$\tilde{N}_{\rm{fbdn}} = \tilde{N} - \tilde{N}_{\rm{alwd}}$
forbidden moves.
We tabulate both allowed and forbidden moves in the same look-up table.
The allowed moves are divided into two groups, one consisting of
acceptable moves which keep the walker within the predefined energy
domain given by Eqs.~(\ref{110}) and (\ref{111}) and the other of
rejectable moves which take the walker outside
this square. If the walker is initially at a configuration
belonging to the innermost square,
all allowed moves are acceptable.
On the other hand, if the walker is not at configuration belonging
to the innermost four energy values, then some allowed moves are
acceptable and some  are rejectable since some moves will take the
random walker outside the square of Eqs.~(\ref{110}) and (\ref{111})
as the largest energy changes in $E$ and  $E_{\rm {I}}$ are
$q\Delta E/2$ and $q\Delta E_{\rm {I}}/2$.

  Whenever the walker visits one of the configuration belonging to
the innermost square we count the number of visit and
take samples of $Q$'s. Next we select a trial move randomly from all
possible $\tilde{N}$ moves. If the selected move is an allowed one,
we update the configuration since it is always acceptable. If the
selected move is forbidden we discard the move but count the current
configuration once more and sample again
for the same configuration to  satisfy the  condition of detailed
balance. This procedure is equivalent to assigning the correct
weight to the sampling configurations. When the walker is at a
configuration not belonging to the innermost four energy values,
we make a random selection of trial move from the group of acceptable
moves only and let the walker move to a new configuration without
wasting computing time. This is possible since we
are not taking data at this configuration.
In this way we can estimate the number of visits for the innermost
four energy configurations, which give relative ratios
$\tilde{\omega}(E^{(i)},E_{\rm{I}}^{(j+1)})/
 \tilde{\omega}(E^{(i)},E_{\rm{I}}^{(j)})$ and
$\tilde{\omega}(E^{(i+1)},E_{\rm{I}}^{(j)})/
\tilde{\omega}(E^{(i)},E_{\rm{I}}^{(j)})$ for all $i$ and $j$.
The sampled data for $Q$ will furnish microcanonical averages
$\tilde{Q}(E,E_{\rm{I}})$.
Since the exact number of configurations $\tilde{\omega}(E,E_{\rm{I}})$
for small values of $E$ and $E_{\rm{I}}$ are easily computed, we can
recursively estimate all $\tilde{\omega}(E,E_{\rm{I}})$'s from these.
To obtain more accurate data we use
a two-step MC simulation. First we accurately estimate
$\tilde{\omega}(E^{(i+1)})/\tilde{\omega}(E^{(i)})$ for all $i$, where
$\tilde{\omega}(E^{(i)}) =
\sum_{E_{\rm {I}}} \tilde{\omega}(E^{(i)},E_{\rm {I}})$.
We then estimate
$\tilde{\omega}(E^{(i)},E_{\rm{I}}^{(j+1)})/
 \tilde{\omega}(E^{(i)},E_{\rm{I}}^{(j)})$ for all $j$.
Now we can easily estimate all $\tilde{\omega}(E,E_{\rm {I}})$ consistently.
The microcanonical average $\tilde{Q}(E^{(i)},E_{\rm{I}}^{(j)})$ is also
estimated directly from data taken during this MC process.
To check the validity of our procedure we compare the estimated
$\tilde{\omega}(E_{\rm{R}},0)$
and $\tilde{\omega}(E_{\rm {I}},E_{\rm{I}})$ for all system sizes with
those of the pure RSOS model and those of the pure Ising model respectively.
We calculated the number of configurations for the pure RSOS model
independently and the those of the pure Ising model are well
known~\cite{kclee}. Both of them agree well with each other to
within expected statistical deviations.

The canonical average of any thermodynamic quantity $Q$ is
obtained by
\begin{equation}
\label{eq.12}
  \left < Q \right > =
	\frac{\sum_{E E_{\rm{I}} M_{\rm{R}} M_{\rm{I}}}
	       Q(E,E_{\rm{I}},M_{\rm{R}}, M_{\rm{I}})
	      \omega(E,E_{\rm{I}},M_{\rm{R}}, M_{\rm{I}})
	      \exp(-CE_{\rm{I}}-AE_{\rm{R}}/2)             }
	     {\sum_{E {E_{\rm{I}}} M_{\rm{R}} M_{\rm{I}}}
	      \omega(E,E_{\rm{I}},M_{\rm{R}},M_{\rm{I}})
	      \exp(-CE_{\rm{I}}-AE_{\rm{R}}/2)             },
\end{equation}
where $M_{\rm{I}}=L^{-2}|\sum_{i}\sigma_{i}|$ is the Ising magnetization,
      $M_{\rm{R}}$ is a RSOS magnetization to be defined in
Eq.(\ref{eq.14}) below and $Q(E,E_{\rm{I}},M_{\rm{R}},M_{\rm{I}})$
and       $\omega(E,E_{\rm{I}},M_{\rm{R}},M_{\rm{I}})$ are the
microcanonical values of the thermodynamic quantity $Q$
and the number of configurations for given
$(E,E_{\rm{I}},M_{\rm{R}},M_{\rm{I}})$, respectively.
In Eq.(\ref{eq.12}), the exponent of the Boltzmann weight factor
does not involve any magnetization. We thus estimate only
$\tilde{\omega}(E,E_{\rm{I}}) \equiv \sum_{M_{\rm{R}} M_{\rm{I}}}
 \omega(E,E_{\rm{I}},M_{\rm{R}},M_{\rm{I}}) $ and
$\tilde{Q}(E,E_{\rm{I}}) \equiv \sum_{M_{\rm{R}} M_{\rm{I}}}
 \omega(E,E_{\rm{I}},M_{\rm{R}},M_{\rm{I}})
      Q(E,E_{\rm{I}},M_{\rm{R}},M_{\rm{I}})/\tilde{\omega}(E,E_{\rm{I}})$.
If $Q$ depends only on $E$ and $E_{\rm{I}}$, as in the case of
the specific heat,
we need only $\tilde{\omega}(E,E_{\rm{I}})$ since
$\tilde{Q}(E,E_{\rm{I}})= Q(E,E_{\rm{I}})$. Otherwise we need to calculate
both $\tilde{\omega}(E,E_{\rm{I}})$ and $\tilde{Q}(E,E_{\rm{I}})$.
The canonical average of
$Q$ can be written as

\begin{equation}
 \left < Q \right > =
    \frac{\sum_{E E_{\rm{I}}} \tilde{Q}(E,E_{\rm{I}})
			      \tilde{\omega}(E,E_{\rm{I}})
			      \exp(-CE_{\rm{I}}-AE_{\rm{R}}/2) }
	 {\sum_{E E_{\rm{I}}} \tilde{\omega}(E,E_{\rm{I}})
			      \exp(-CE_{\rm{I}}-AE_{\rm{R}}/2) }.
\label{2}
\end{equation}
All thermodynamic quantities we use in this paper depend only on the
energies $E$ and $E_{\rm{I}}$ or only on magnetizations $M_{\rm{R}}$ and
$M_{\rm{I}}$.
For example, $Q(E,E_{\rm{I}},M_{\rm{R}},M_{\rm{I}}) = E$ for the total
energy, $Q = M_{\rm{I}}$
for the Ising magnetization and
$Q = \delta_{M_{\rm{I}},M}$ for the magnetization probability distribution
$P_L(M)$.
In all cases, samplings are performed in the 2D plane of $(E,E_{\rm{I}})$.
The number of possible $E$ and $E_{\rm{I}}$ is proportional to $L^2$
and the simulation time increases as $L^{4}$ where $L$ is the lattice size.
This makes it difficult to simulate systems of large size. Nevertheless,
once $\tilde{Q}(E,E_{\rm{I}})$ and
$\tilde{\omega}(E,E_{\rm{I}})$ are
estimated, the canonical average $\left < Q\right >$ can be easily
calculated for all $A$ and $C$ from Eq.~(\ref{2}).

As the specific heat of the 2D $XY$ model has no diverging peak as
$L\rightarrow\infty$, the RSOS part of the specific heat may also not
diverge which
implies the transition temperature may not be easy to determine accurately.
For the 2D $XY$ model, which has vanishing magnetization in the bulk
limit for any finite temperature, the helicity modulus, or stiffness
constant which is a measure of the quasi long-range order of the order
parameter describing the continuous $U(1)$ symmetry and the superfluid density
of a Bose superfluid~\cite{fisher}, may be used to determine
the $XY$ transition temperature from the finite size scaling form~\cite{weber}
at the KT point. The analogous quantity in the RSOS model is the free energy
of a step in the height which has the finite size scaling form~\cite{blnigh}
$F_{s}(L)=\pi /4+1/(A+BlnL)$ at the roughening temperature. However, the
computation of this requires the partition function with two different
boundary conditions which is time consuming. As an alternative method,
we know that the roughening transition in the RSOS model is
one between flat and rough phases. We introduce an
order parameter, analogous to the  magnetization of an Ising model,
$M_{\rm{R}}$, which measures the flatness of the RSOS
system, as
\begin{equation}
\label{eq.14}
  M_{\rm{R}} = L^{-2} \sum_I m_{\rm{R}}(h_I),
\end{equation}
where $m_{\rm{R}}(h_I)=1$ if the height $h_I$ is equal to
the majority height and $m_{\rm{R}}=0$ otherwise. It is clear from this
definition
that $M_{\rm{R}}=1$ for a completely flat phase  and vanishes in a rough
phase in the thermodynamic limit.
In Figs.~2(a) and 2(b) we plot $M_{\rm{R}}$ and $M_{\rm{I}}$ in
the parameter space
$(A,C)$ for lattice size $L=22$. It is easily seen from the figures that
there are three different phases depending on the parameters $A$ and $C$:
Ising ordered rough (IOR), Ising ordered flat (IOF) and Ising disordered
flat (DOF). The IOR and
DOF phases are separated by the IOF phase in the low temperature region
though it is not clear whether, in the region of small $A$ and $C$, the two
phases are also separated or not. The corresponding phases of the $XY$-Ising
model of Eq.(\ref{0}) are $XY$ and Ising order (IOR), $XY$ disorder and Ising
order (IOF) and $XY$ and Ising disorder (DOF). It is interesting to note
that the fourth phase of $XY$ order and Ising disorder, corresponding to an
Ising disordered rough phase of the RSOS-Ising model, does not exist.

We now examine the behavior of the specific heat.
For simplicity, we consider $C_{\rm{R}}$ and $C_{\rm{I}}$
corresponding to the RSOS and Ising specific heats, respectively.
These are defined as
$L^2C_{\rm{S}} = -D^2 \partial \left < E_{\rm{S}} \right > /
			\partial D,$
where $D$,S$=A$,R and $C$,I respectively.
In Figs.~3(a) and 3(b) we plot the RSOS and Ising specific heats
for lattice size $L=22$.
We have obtained the specific heats for several lattice sizes up to $L=22$.
The RSOS specific heat ridge parallel to the $C$-axis ceases to grow with
increasing $L$ for $L\geq 18$ as for the standard $XY$ model while the
Ising specific heat ridge continues to grow as the system size
increases. From the specific heats of several lattice sizes
we find that the growth rate of the specific heat ridges in the
small $A$ and $C$ region is no higher than that of the Ising ridge
parallel to $A$-axis which implies that the exponent $\alpha\leq 0$
for either contribution. Because of the slow increase of the Ising ridge
with $L$ and the limitation to fairly small $L$, we are unable to obtain any
reasonable estimate of the exponent $\alpha/\nu$.

Having established the existence of roughening and Ising
transitions we can now locate the precise location of the two transition
lines. It is relatively simple to locate the Ising transition line by
using the probability distribution of the Ising magnetization $M_{\rm{I}}$,
$P_L(M_{\rm{I}},A,C)$.
In a phase with Ising order,
$P_L(M_{\rm{I}}=0,A,C)$ decreases to zero as the system
size increases while $P_L(M_{\rm{I}}=0,A,C)$ increases as $L$ increases in
an Ising disordered phase. We then expect that there will be a unique
crossing point (A*,C*) in a plot of
$P_L(M_{\rm{I}}=0,A,C)$ curves for several lattice sizes along a path in
the $A$-$C$ plane if the path crosses the transition line. Note that this
does not use the concept of finite size scaling.
This probability distribution method is a powerful tool to locate the
transition point if the system is appropriate to this type
of analysis.
We have obtained $P_L(M_{\rm{I}}=0,A,C)$ curves in the $A$-$C$ plane for
lattice sizes $L=10, 14, 18, 22$ and found that there exists a
well defined line of intersection points from several
$P_L(M_{\rm{I}}=0,A,C)$ surfaces which is shown in Fig.~4(b).
The estimated positional uncertainty of the Ising transition line
in the $A$-$C$ plane is at most 0.001. Unfortunately, it is difficult to
find any RSOS order parameter suitable for this type of analysis to locate
the transition line and,
as an alternative method, we use the position of the peak in the
susceptibility  of the RSOS order parameter $M_{\rm{R}}$. In the rough
phase, this susceptibility is infinite in the thermodynamic limit and, for
finite $L$, the susceptibility peak grows as $L^{2-\eta}$.
In the thermodynamic limit, the position of this
diverging peak gives the transition point.
The peak position in a finite system, for fixed $C$, approaches its
bulk value $A^{*}$ as
\begin{equation}
  \left | A_{\rm{peak}}(L) - A^{*}\right | \propto L^{-\frac{1}{\nu}},
\label{3}
\end{equation}
where $A_{\rm{peak}}(L)$ is the peak position for lattice size $L$.
In Fig.~4(a) we plot $A_{\rm{peak}}(L)$
versus $L^{-1}$ for several $C=0.10, 0.15, 0.20,$ and 0.25 from bottom
to top. All points are on almost straight lines which implies that
Eq.~(\ref{3}) describes the scaling relation of the peak position with
$\nu=1$. Extrapolating to $L^{-1}=0$,
we obtain the bulk value of $A^{*}$ for several fixed $C$'s.
The specific heat exponent $\alpha\leq 0$ along the
roughening transition line, as we argued earlier,
and this implies $\nu \geq 1$. In fact, at the roughening transition
point which is the KT point in the $XY$ model, the exponent $\nu =\infty$.
If $\nu >1$,  $A_{\rm{peak}}$ is convex up in $L^{-1}$ and
the extrapolated $A^{*}$ under
the assumption $\nu=1$ is an upper bound for the transition
point. In Fig.~4(b) we draw the roughening transition line extrapolated from
Eq.(\ref{3}) with $\nu =1$ together with the Ising transition line.
The IOR and DOF phases are completely separated by the IOF phase in all
parameter space $(A,C)$. Since the phases are separated when we use an upper
bound for $A^{*}$, they will certainly be separate for the true values of
$A^{*}$ as this will increase the separation.
The phase trajectory of the FF$XY$ model goes from IOR to DOF as the
temperature increases from zero and the result is strong evidence
for the double transition scenario in the FF$XY$ model.

Since the two phase transition lines  are accurately located some critical
parameters along these lines can be estimated.
We have calculated the Ising susceptibility from the usual definition
and estimated the exponent $\gamma/\nu$
along the Ising transition line using sizes $L=6,10,14,18,22$ for the fit
(See Fig.~5). The exponent
$\gamma/\nu$ is consistent with the Ising value down to $A=0.8$ but deviates
significantly for smaller $A$ where the two transition lines become close.
Presumably this discrepancy is due to large corrections to scaling due to
large fluctuations of the RSOS order and the limited system sizes.
The height-height correlation function of the RSOS model defined by
$G({\bf r}_I - {\bf r}_J) \equiv \left < (h_I - h_J)^2 \right >$
behaves as
$ G({\bf r}_I - {\bf r}_J) \sim a(A,C) \log(\left | {\bf r}_I-{\bf r}_J
\right |)$ in the rough phase.
The coefficient $a(A,C)$ changes continuously in the rough phase and has
a universal value $2/\pi^{2}$ at the roughening transition
point.~\cite{nienhuis} The height width $\sigma^2$ is defined by
$\sum_I(h_I-\bar{h})^2/L^2$, where $\bar{h}$ is the average height and,
by integrating the height-height
correlation function,
$\sigma^2 \approx \frac{a}{2} \log(L) + b$ in the rough phase.
From the MC simulations, we have obtained $\sigma^2(A,C)$ in the IOR phase
and found that the height width $\sigma^{2}$ agrees well with the $\log L$
scaling
form in the whole IOR region. The coefficient $a$ increases as $A$ decreases
from $A^{*}$ into the rough phase, as expected.
Along the roughening transition line where two transitions are not
close to each other, the coefficient $a\approx 0.16$, which is somewhat
smaller than the universal value $2/\pi^{2}\approx 0.203$.
We find that the coefficient $a \approx 0.19$
along the roughening transition line where two transitions are very
close to each other.
We do not use any scaling or extrapolation method to locate
the RSOS transition points in the region where two transitions are well
separated but simply use the RSOS susceptibility peak
positions of the largest lattice size, $L=22$, as the transition points.
This peak position moves very slowly toward smaller $A$ as the system
size becomes larger, as expected.
In fact the estimated $a$ along the roughening transition line in that
region  decreases slightly as the system size increases. This means that
the estimated transition line actually lies in the flat region
close to the real roughening
transition line which will lie to the left of our estimated line.
We have located a more reasonable  roughening
transition line
in that region using the criterion $a = 2/\pi^{2}$ and find that the
transition line shifts to the left about $6\%$ in $A$. This is consistent
with our arguments that the true $A^{*}$ is smaller than the value estimated
on the assumption that $\nu =1$ at the roughening transition and that the
two transitions are separated by a thin tongue of IOF phase.

\section{Conclusions}

We have investigated the phase transitions in a restricted solid on solid
model coupled to an Ising model by a rejection free microcanonical Monte
Carlo method. This model is derived from the coupled $XY$-Ising system by a
duality transformation on the $XY$ degrees of freedom and by making what
appear to be reasonable restrictions on the allowed height differences. The
$XY$-Ising model is a coarse-grained generalization of the fully frustrated $XY$
model
which was originally introduced in the hope that it would make numerical
studies of the FF$XY$ system easier. Despite this convoluted chain of
transformations and approximations, the RSOS-Ising model is believed to
undergo continuous phase transitions in the same universality classes as the
FF$XY$ system. We believe that the restriction on the heights $|\Delta h|=0,1$
is not a serious one in that it will not affect the universality class of
the transitions provided they are continuous and that the topology of the
phase diagram is also not qualitatively affected. We believe that the only
possible difference lifting the RSOS restriction on the heights could have
would be to allow the Ising and roughening lines to merge into a segment of
first order transition line separating the DOF and IOR phases. In our
numerical study of the RSOS-Ising system we identified three phases: Ising
ordered rough (IOR), Ising ordered flat (IOF) and Ising disordered
flat (DOF) which correspond to $XY$ and Ising ordered, $XY$ disorder and Ising
order and complete disorder respectively. Using the rejection free
microcanonical method, we obtain a precise phase diagram and demonstrate
that the IOR and DOF phases are separated by a sliver of IOF phase for all
physical values of the parameters. As the temperature in the FF$XY$ model is
raised, the system will follow some path in the $A,C$ plane from the IOR
phase to the DOF phase. Since it must pass through the intervening IOF
phase, this is strong evidence that there are two continuous transitions in
the FF$XY$ and $XY$-Ising models.

The critical exponents along the Ising transition line seem to change
continuously in the region where the Ising and roughening transitions are
close together but this may be due to large corrections to scaling
caused by the nearby roughening transition.
In that case the correct determination of the critical behavior will require
very large systems. Since we have located the precise phase transition
lines the true critical behavior
along the Ising line may be determined using other MC methods appropriate
for the very large systems needed.
We leave this for future study.

\acknowledgements
This work was supported in part by the NSF through Grant No.
DMR-9222812 (JMK), in part by the Ministry of Education, Republic
of Korea through a grant to the Research Institute for Basic Sciences, Seoul
National University, in part by the Korea Science Foundation through
Research Grant to the Center for Theoretical Physics, Seoul National  
University, and in part by SNUCTP-Brown exchange program. SL would like to
thank the DaeWoo Foundation for financial support. KCL would like to thank 
Kyungsik Kang and many people of Brown University for the hospitality
extended to him during his stay at Brown University on his last sabbatical
leave.

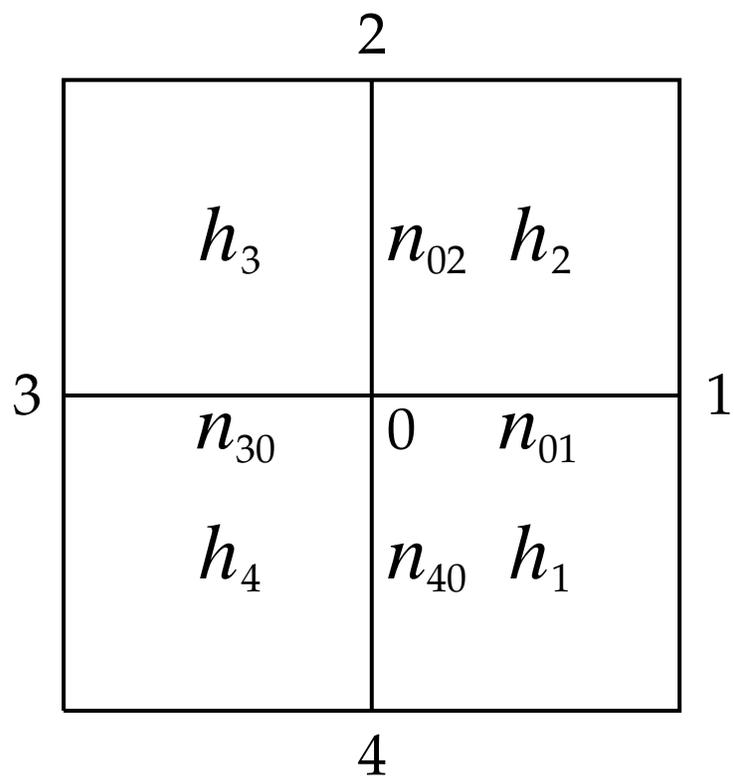
\begin{figure} 
\caption{Representation of bond variables by integer height variables on
the dual lattice. Under the height variable scheme all constraints on
bond variables are automatically satisfied. }
\end{figure} 
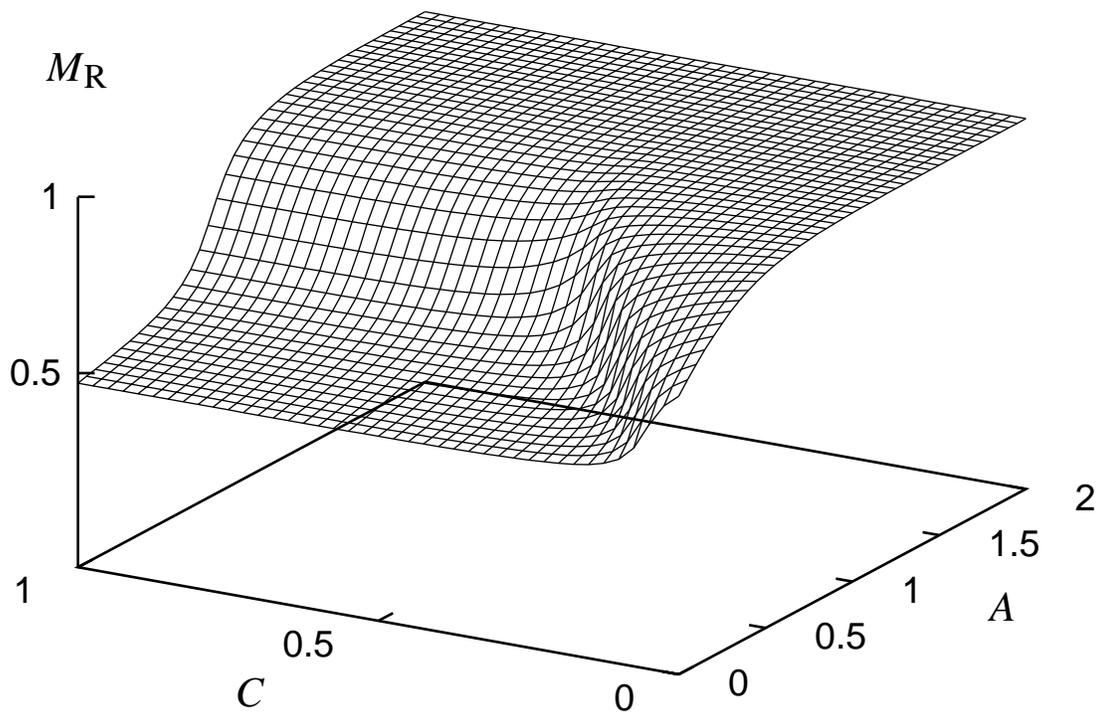
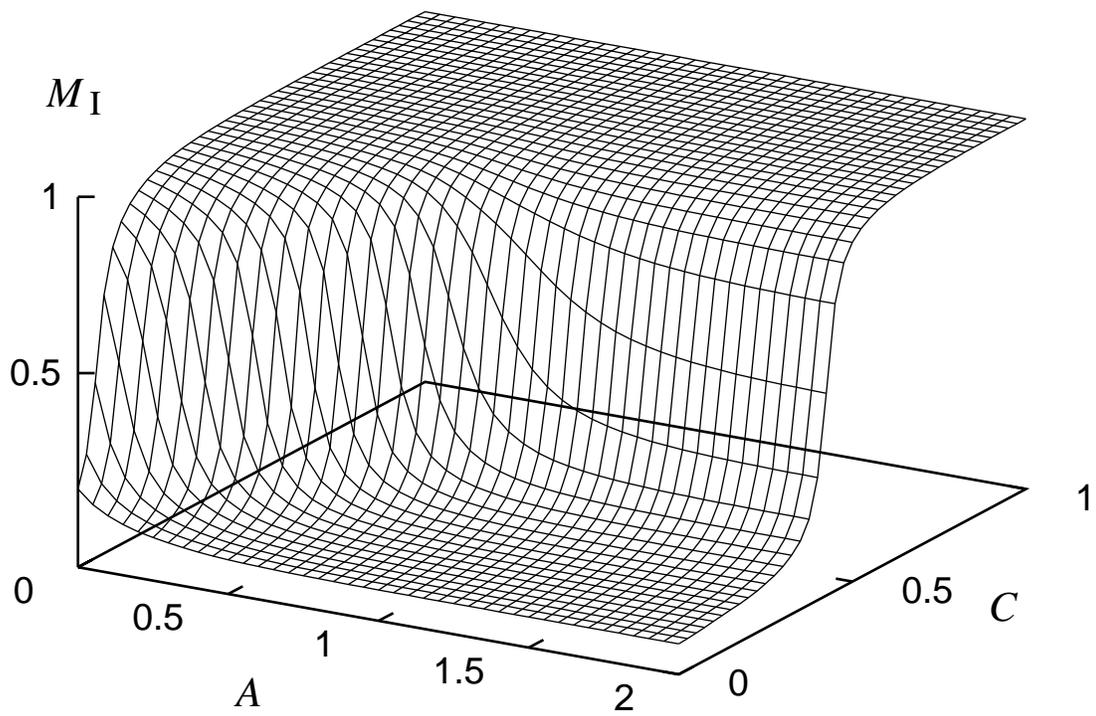
\begin{figure}
\caption{(a) Surface flatness $M_{\rm{R}}$ of the RSOS part for lattice size
$L=22$. It is easily seen that there is a roughening transition 
between flat($M_{\rm{R}} \simeq 1$) and rough phases($M_{\rm{R}} \simeq 0.5$). 
(b) Magnetization $M_{\rm{I}}$ of the Ising 
part for lattice size $L=22$. There is a relatively sharp transition
region in the Ising magnetization. In the region of small $A$ and $C$
the Ising transition line is close to the roughening transition line. 
It should be noted that the viewing angle of (b) is $90^{\circ}$ rotated
from that of (a).}
\end{figure} 
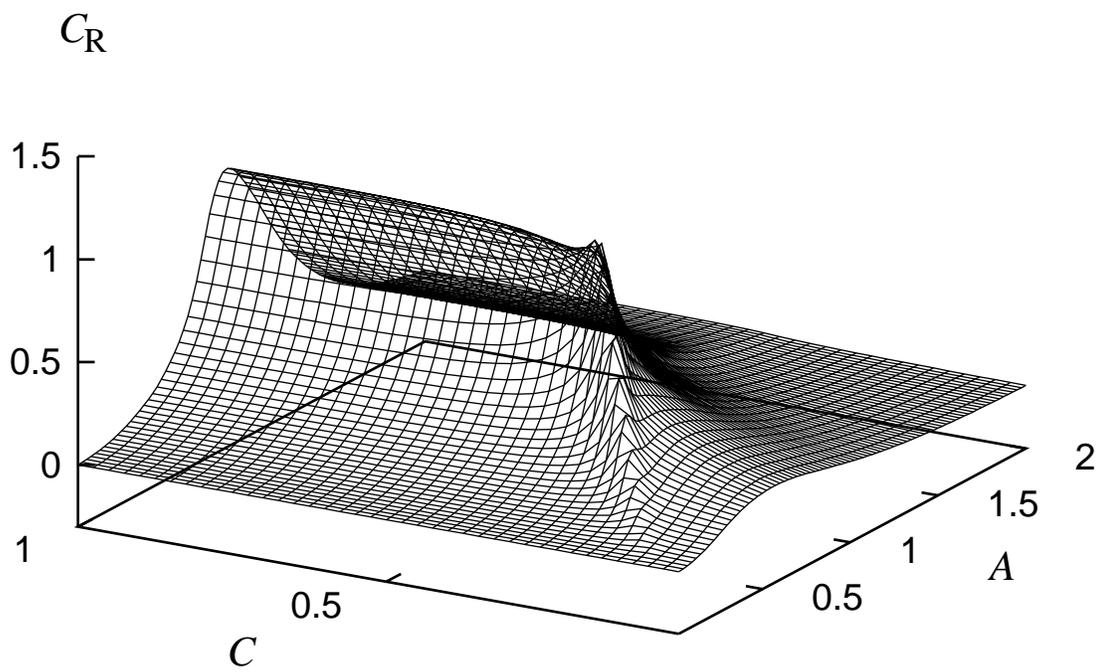
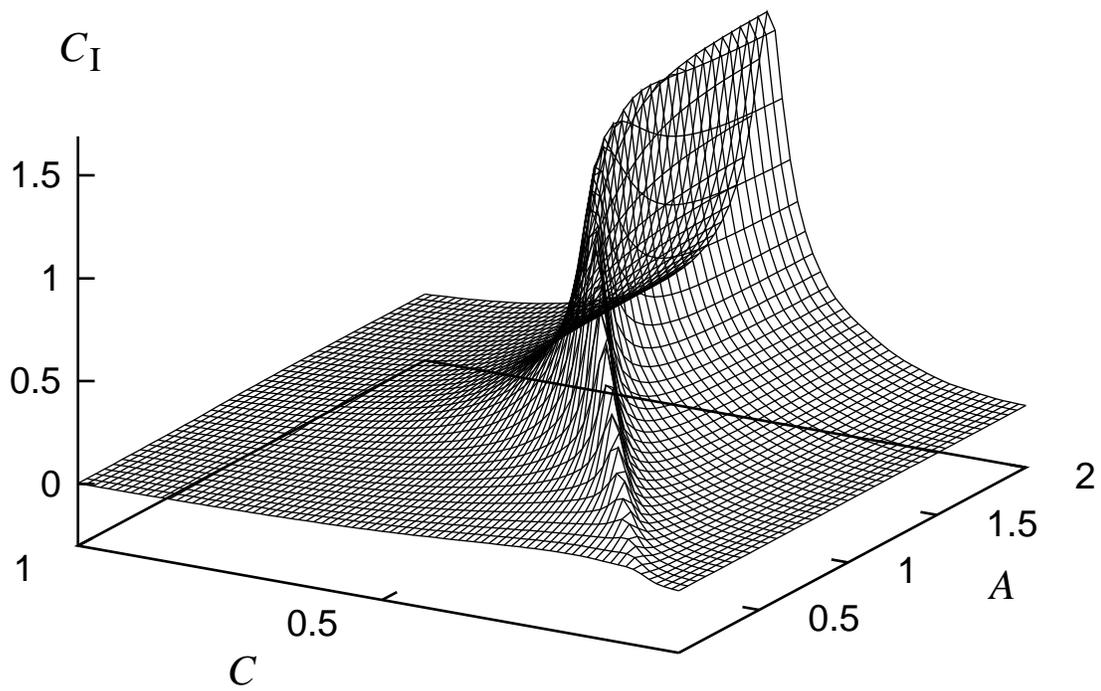
\begin{figure}
\caption{ Specific heat $C_{\rm{R}}$ of the RSOS part for lattice size $L=22$.
The specific heat ridge parallel to $C$-axis does not increase
for $L\geq 18$ just as in the $XY$ model.
(b) Specific heat $C_{\rm{I}}$ of the Ising 
part for lattice size $L=22$. As expected, the specific heat ridge grows as
$L$ increases. The location of the Ising ridge is very close to
that of the RSOS ridge in small $A$ and $C$ region.}
\end{figure}
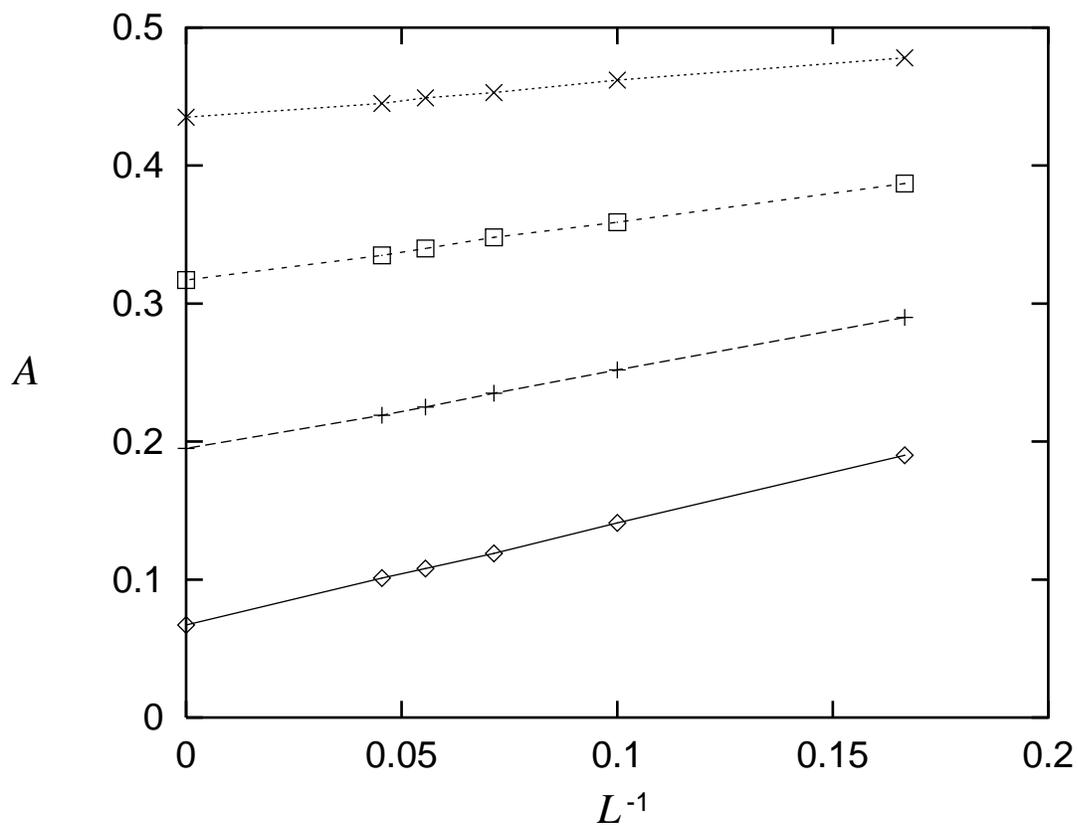
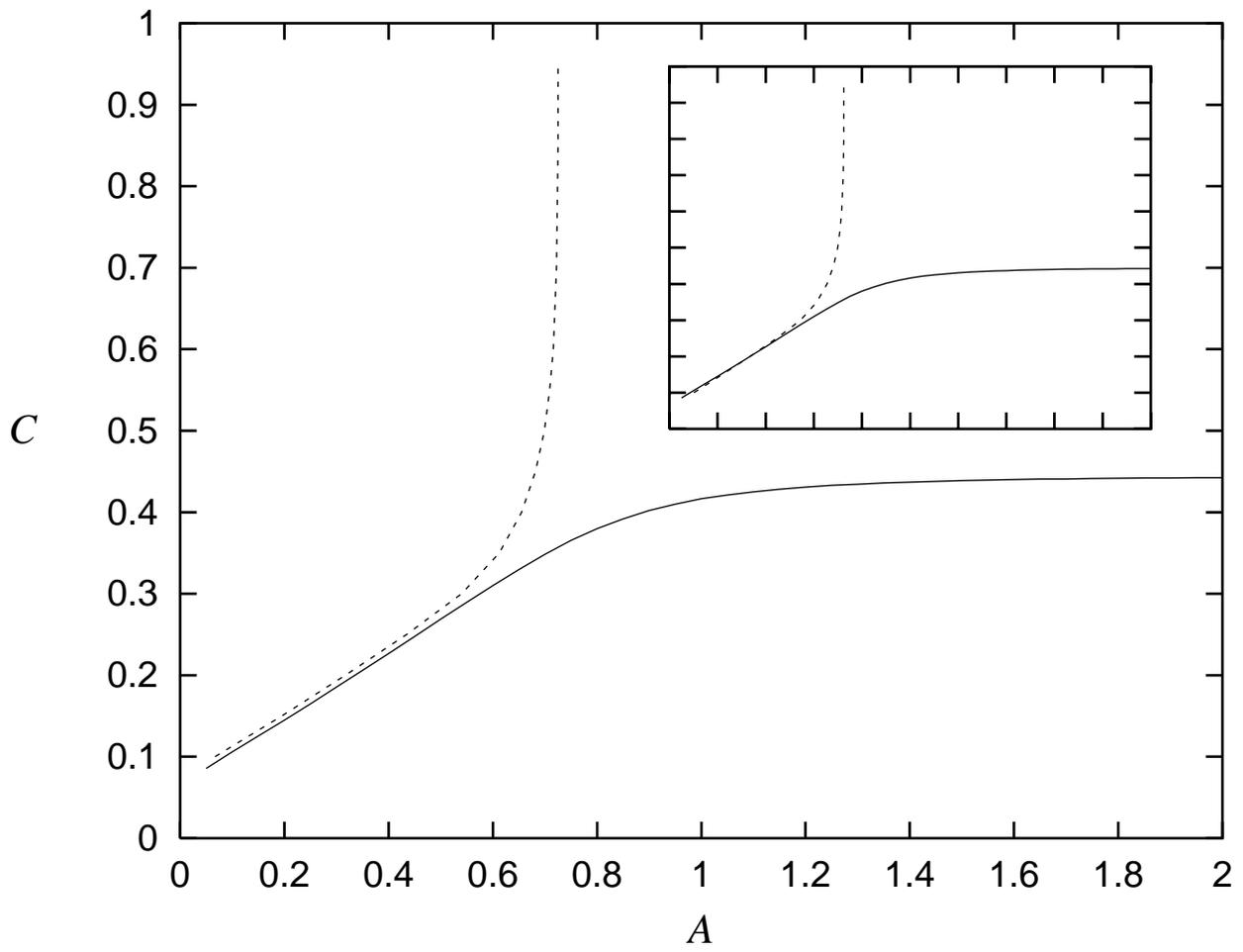
\begin{figure}
\caption{ (a) Extrapolating the peak position of the $M_{\rm{R}}$ susceptibility
in the bulk limit for several $C$ values. Each line corresponds to 
$C=0.10, 0.15, 0.20$ and 0.25 from bottom to top. The data points on the
$A$-axis ($\frac{1}{L} = 0$) indicate extrapolated values.
(b) Phase diagram of the RSOS-Ising model in the $A,C$ plane. The solid
and dotted lines represent the Ising and roughening transition, respectively.
The roughening transition line is obtained by extrapolations  of the 
$M_{\rm{R}}$ susceptibility peak positions. For the comparison we present another
phase diagram in the inset, where the roughening transition line
is the position of peak in the $M_{\rm{R}}$ susceptibility for $L=22$.}
\end{figure}
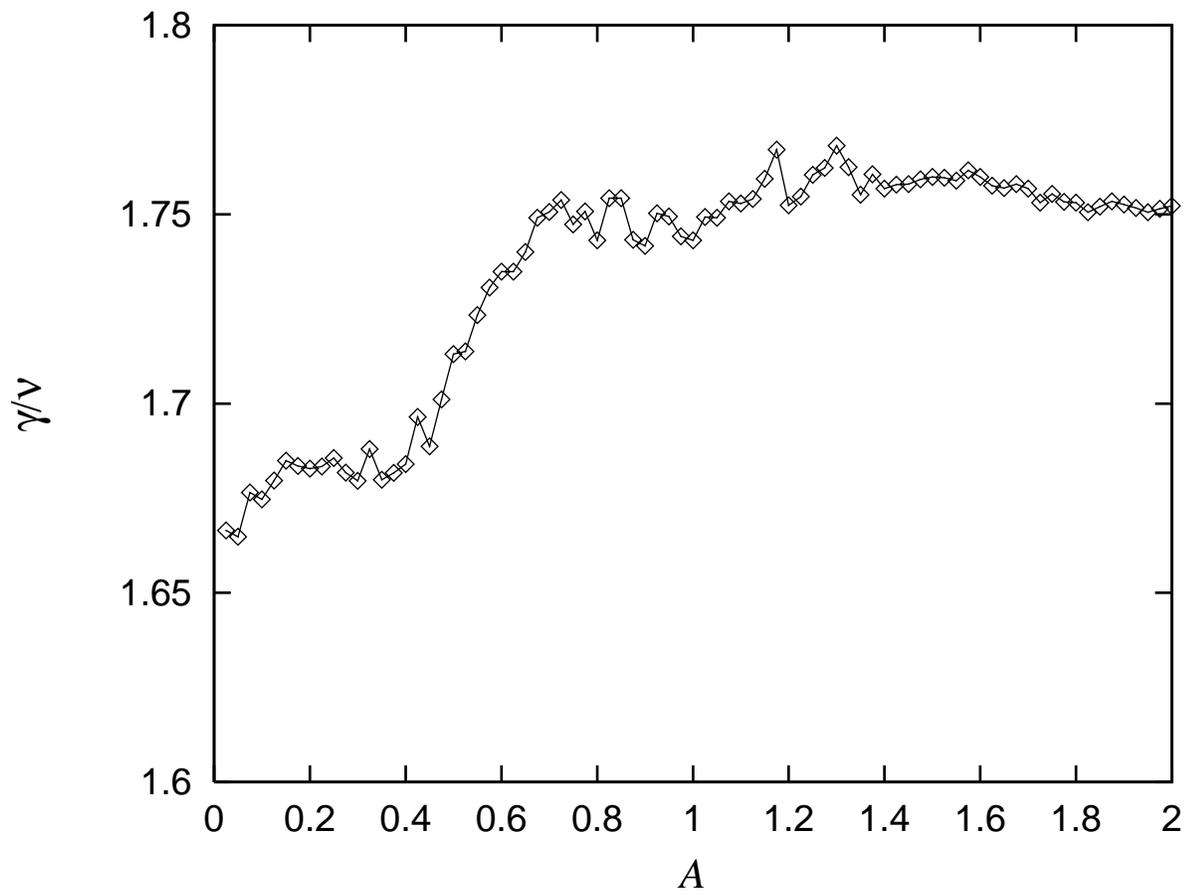
\begin{figure}
\caption{ Exponent $\gamma/\nu$ of the Ising susceptibility along the Ising
transition line. For $A>0.8$ $\gamma/\nu\approx 1.75$ as in the 2D Ising
model. This value is obtained by using the best fit for the finite-size scaling
relation for the susceptibility $\chi \propto L^{\gamma/\nu}$ with $L$=
6, 10, 14, 18, and 22.} 
\end{figure} 
\end{document}